\documentclass[prd,showpacs,letterpaper,twocolumn]{revtex4}%
\usepackage{graphicx}
\usepackage{bm}
\usepackage{epsf}
\usepackage{rotating}
\usepackage{epsfig,graphics,rotate,color}
\usepackage{wrapfig}
\usepackage{amssymb}
\usepackage{amsmath}
\usepackage{amsfonts}
\usepackage{array,hhline,dcolumn}%
\providecommand{\U}[1]{\protect\rule{.1in}{.1in}}
\begin{document}
\title{Electron Antineutrino Disappearance at KamLAND and JUNO as Decisive
  Tests of the Short Baseline $\bar \nu_\mu \rightarrow \bar \nu_e$ Appearance Anomaly}
\author{J.M. Conrad$^{1}$ and M.H. Shaevitz$^{2}$}
\affiliation{$^{1}$ Massachusetts Institute of Technology}
\affiliation{$^{2}$ Columbia University}

\begin{abstract}

The IsoDAR antineutrino source, which produces a flux from $^8$Li
isotope decay at rest, when paired with the 
proposed JUNO (Jiangmen Underground Neutrino Observatory) detector, has
 unprecedented sensitivity to $\bar \nu_e$ disappearance for
 oscillations at high $\Delta m^2$.   Assuming
 $CPT$ invariance,  the sensitive region for $\bar \nu_e$ 
 disappearance can be used to restrict the allowed
 parameter space of a $\bar \nu_\mu \rightarrow \bar \nu_e$ appearance signal.
 The 5$\sigma$ sensitivity of this experiment
 covers the entire
 $\bar \nu_\mu \rightarrow \bar \nu_e$ allowed parameter space from a
 combined fit to short-baseline experiments.  This represents a decisive test of the 
 LSND and MiniBooNE antineutrino appearance signals within all models that are $CPT$ invariant.
 Running IsoDAR at KamLAND restricts a large part of the appearance signal region in a similar way.
  
\end{abstract}

\maketitle

\section{Introduction}

Within a model with $CPT$ invariance, the allowed parameter space for $\bar
\nu_\mu \rightarrow \bar \nu_e$ appearance oscillations must be contained within
the parameter space allowed for $\bar \nu_e$ disappearance, as can be seen
from the following chain of reasoning:
\begin{enumerate}
\item  $CPT$ invariance requires that $\bar \nu_\mu \rightarrow \bar \nu_e$ oscillations
  and $\nu_e \rightarrow \nu_\mu$ oscillations must be identical.
\item    The probability for $\nu_e$ disappearance must be larger than the probability
 for $\nu_e \rightarrow \nu_\mu$ oscillations.  
\item $CPT$ invariance requires that the probability of $\bar \nu_e$ disappearance be the
same as the probability for $\nu_e$ disappearance.   
\end{enumerate}
Thus,  if the sensitive region of a $\bar\nu_e$ disappearance experiment
entirely covers the parameter space for a $\bar \nu_\mu \rightarrow
\bar \nu_e$ signal region,  then either a signal must be observed with mixing
angle such that
$\sin^2 2\theta_{\bar{ee}} > \sin^2 2\theta_{\bar{\mu e}}$, or all models based on
oscillations that assume $CPT$ invariance must be ruled out as an
explanation for the $\bar \nu_\mu \rightarrow
\bar \nu_e$ appearance signal.  We will use this argument to show how the
JUNO  (Jiangmen Underground Neutrino Observatory) and KamLAND
detectors can be used to address existing short-baseline $\bar \nu_\mu
\rightarrow \bar \nu_e$ appearance signals.

  \begin{figure}[t]\begin{center}
{\includegraphics[width=3.in]{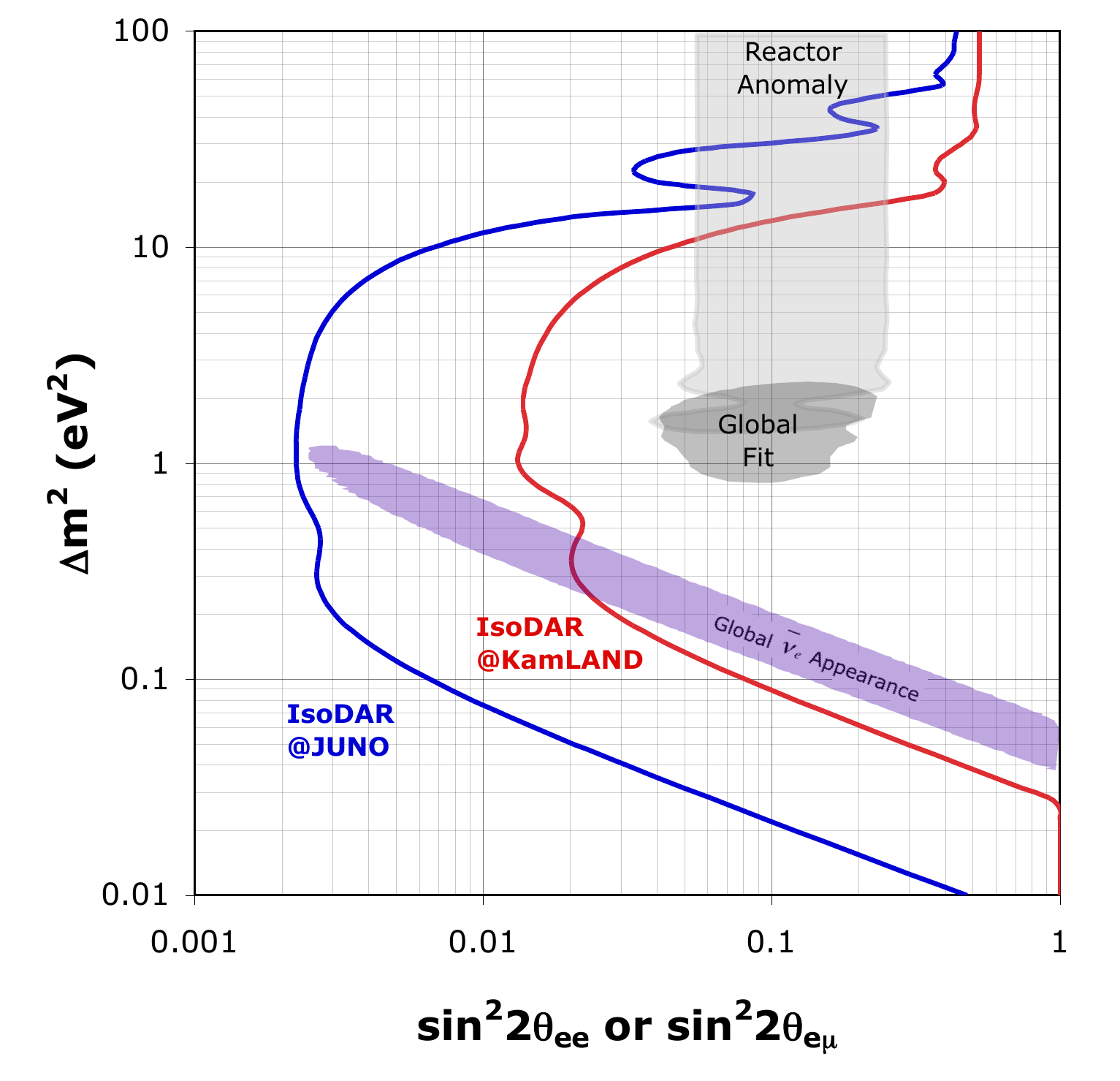}
} 
\end{center}
\vspace{-0.25in}
\caption{Allowed regions and sensitivities for various experiments.    
The red and blue solid curves indicate $\Delta m^2$ vs. $\sin^2 2\theta_{ee}$ 
boundaries where the
null oscillation hypothesis can be excluded at 5$\sigma$ with IsoDAR@KamLAND
and IsoDAR@JUNO experiments respectively for five-year data runs.  
Also, shown by the light (dark)
gray areas are the 99\% allowed regions for the Reactor Anomaly\cite{reactor} 
(Global Oscillation Fit\cite{Giunti_pragmatic}).  Finally, the purple region corresponds
to the $\Delta m^2$ vs. $\sin^2 2\theta_{e\mu}$ allowed region at 99\% CL
from a combined fit to all  $\bar\nu_e$ appearance data\cite{sbl}.
\label{JUNO_sense} }
\end{figure}

The world's data on $\bar \nu_\mu \rightarrow \bar \nu_e$ appearance
from short-baseline experiments can be combined to yield a ``Global
$\bar \nu_e$ Appearance''
allowed region, using the code from Ref.~\cite{sbl}.   Three
experiments contribute.
The LSND experiment observed a $>3\sigma$ excess of $\bar \nu_e$ in a
$\bar \nu_\mu$ beam that can be interpreted as  $\bar \nu_\mu \rightarrow
\bar \nu_e$ signal at the mass splitting of $\Delta m^2>0.01$ eV$^2$ \cite{LSND}.
The complementary KARMEN experiment failed to observe $\bar \nu_\mu \rightarrow
\bar \nu_e$ oscillations \cite{KARMEN}, and this represents an
important constraint.     These studies were
followed by the MiniBooNE experiment, which employed both $\nu_\mu$ and
$\bar \nu_\mu$ beams.    For this discussion, we consider only the
antineutrino running, which can be directly compared to KARMEN and
LSND with no assumptions concerning $CP$ violation.
MiniBooNE observed an excess of $\bar \nu_e$ events \cite{MBnubar}. 
Fig.~\ref{JUNO_sense} shows the 99\% confidence level (CL)
allowed region for appearance from a joint fit to the data sets (purple), where 
$\Delta m^2$ is plotted versus $\sin^2 2\theta_{\mu e}$.     

Also relevant to this discussion are recent analyses of
$\bar \nu_e$ disappearance signals from reactors \cite{reactor} 
that have reported a signal at $>2 \sigma$.   The light gray
region of Fig.~\ref{JUNO_sense}  shows the 99\%  allowed region for this disappearance
in terms of  $\Delta m^2$ versus $\sin^2 2\theta_{e e}$ from
Ref.~\cite{reactor}.     This is called the ``Reactor Anomaly.''

These two results are inconsistent
with a three-neutrino model \cite{Sorel}, but can be explained if one or more
noninteracting (``sterile'')  neutrinos are introduced \cite{sbl, Sorel, 
  Giunti_pragmatic}.     Other data which are included in the
sterile neutrino global fits include low-level signals for $\nu_e$ appearance from sources
\cite{SAGE3, GALLEX3}, constraints in $\nu_e$ oscillations from
comparisons of KARMEN and LSND cross-section measurements
\cite{ConradShaevitz}, and a large number of null signals from
$\nu_\mu$ oscillation experiments \cite{sbl}.
The $\Delta m^2$ versus $\sin^2 2\theta_{e e}$
allowed region at 99\% CL for the $\bar \nu_e / \nu_e$
disappearance parameters from a combined fit to all oscillation data is
indicated by the dark gray region \cite{Giunti_pragmatic}, which we will call the 
``Global Fit''  99\% CL allowed region.    
  
\section{IsoDAR@KamLAND}

The IsoDAR@KamLAND experiment \cite{PRL} is proposed to decisively address the 
Global Fit.       This experiment pairs the IsoDAR source with
the KamLAND detector.    IsoDAR makes use of the same cyclotron
design as the injector cyclotron for DAE$\delta$ALUS to accelerate
protons to 60 MeV/n.      
The protons impinge on a Beryllium target that produces copious neutrons.
The target is 
surrounded by a 99.99\% isotopically pure $^7$Li sleeve, where neutron
capture results in $^8$Li production.  The $^8$Li isotopes then undergo
$\beta$ decay at rest to produce an isotropic
$\bar \nu_e$ flux with an average energy of $\sim$6.5 MeV and
an endpoint of $\sim$13 MeV.    

In a liquid scintillator detector,  events are observed through
inverse beta decay (IBD), $\bar \nu_e +p \rightarrow e^+ + n$,
which is easily tagged through
positron (prompt-light) --neutron-capture coincidence.   The energy of the
neutrino can be reconstructed from the visible energy of the positron:  
$E_\nu = E_{e^+} + 0.8 \ {\rm MeV } $. 
For KamLAND, the energy
resolution of $6.4\%/\sqrt{E (\rm MeV)}$ is assumed in the sensitivity
calculations \cite{PRL}.
The position can be
reconstructed using the timing of arrival of the scintillation light
at the photomultiplier tubes (PMTs).   Thus, this interaction allows for accurate
reconstruction of $L/E$.   The vertex resolution for KamLAND is 
assumed to be $12 {\rm cm}/\sqrt{E  (\rm MeV)}$ \cite{PRL}; however, 
the resolution is small compared to the variation of the extent of the 
neutrino source, which leads to an uncertainty in the neutrino
flight path of 40 cm.

The analysis to obtain the oscillation sensitivity follows the
method of Ref.~\cite{sblarticle} assuming a 5\% normalization uncertainty
and a 90\% detection efficiency.  Since L and E can be precisely 
measured, this analysis exploits the L/E dependence of the possible
oscillation probability, $P = 1 - \sin^2 2\theta \sin^2[1.27\Delta m^2(L/E)]$,
to estimate the $\Delta m^2-\sin^22\theta$ regions where the null oscillation
hypothesis can be excluded at the 5$\sigma$ confidence level.

In less than one year of running this
experiment, using the 1 kton KamLAND liquid scintillator detector, 
will indicate a signal if the Global Fit is due to
oscillations,  and with five years of running, will achieve the sensitivity at
5$\sigma$ shown in Fig.~\ref{JUNO_sense} (red line).     This result directly addresses
the reactor disappearance signal up to $\Delta m^2 \sim 10$ eV$^2$.
(The higher $\Delta m^2$ region is not accessible unless IsoDAR
normalization, which is presently assumed to be 5\% \cite{PRL}, can be greatly 
improved.)   

This experiment also makes strong statements concerning the sterile
neutrino model for the Global Fit (dark gray region).
If no signal is
observed,  then the sterile neutrino model for this region is decisively ruled out.
The value of running
for five years or more, if an oscillation signal is observed, is that the L/E
pattern can be mapped out to determine the number of sterile
neutrinos involved in the oscillation \cite{PRL}.

  \begin{figure}[b]\begin{center}
{\includegraphics[width=3.in]{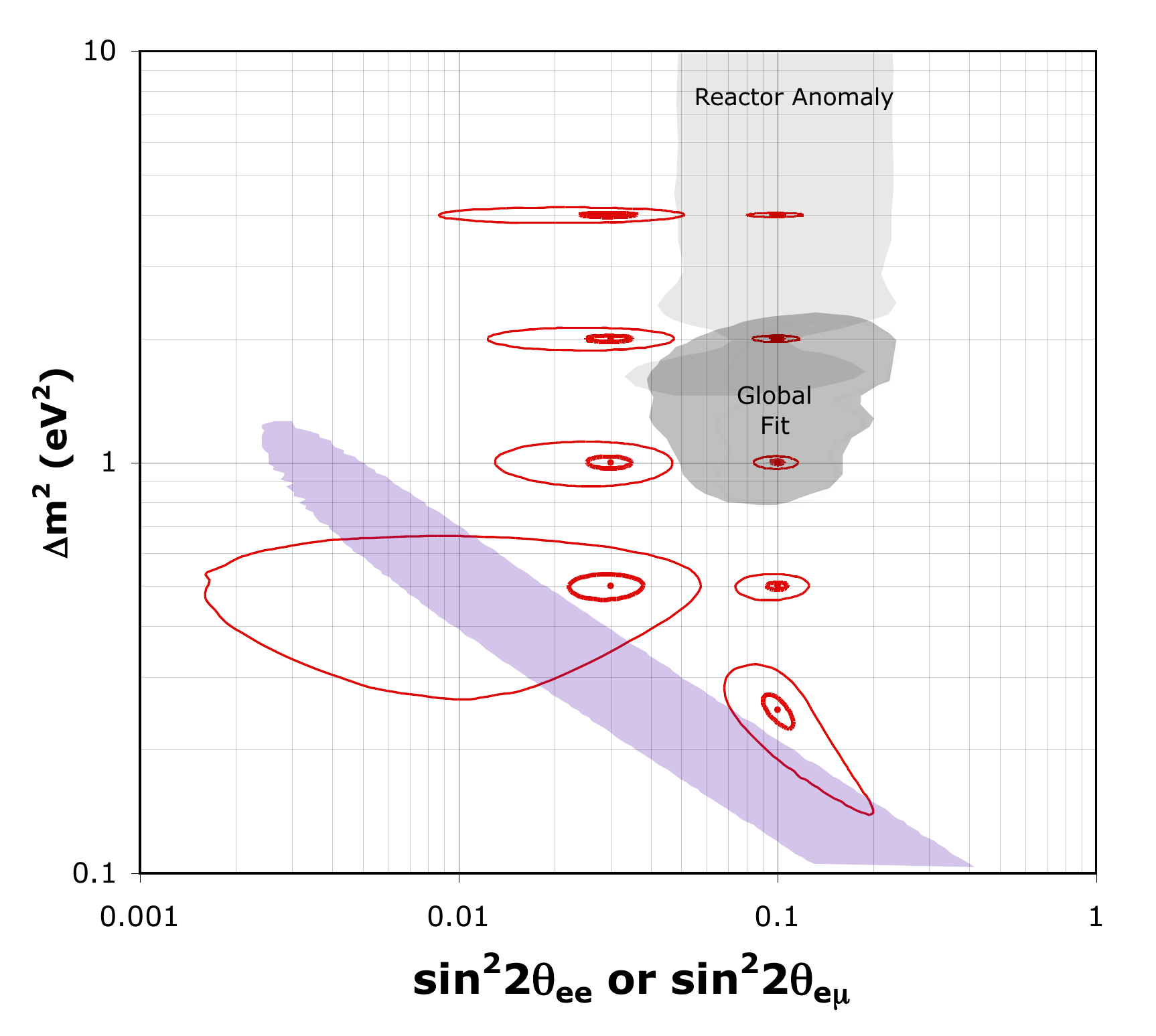}
} 
\end{center}
\vspace{-0.25in}
\caption{Measurement capability of IsoDAR@KamLAND for five years of
  running overlaid on the  Reactor Anomaly (light gray) and
  Global Fit (dark gray) $\Delta m^2$-$\sin^22\theta_{ee}$
  and the Global $\bar \nu_e$ Appearance (purple) $\Delta m^2$-$\sin^22\theta_{\mu e}$
  allowed regions.  1 and 5$\sigma$ measurement contours are indicated
  for various underlying true sets of parameters.
\label{measure} }
\end{figure}

New in this report, we point out that 
the IsoDAR@KamLAND 5$\sigma$ sensitivity also covers 
a substantial region of the $\bar \nu_e$ appearance anomaly. 
As shown in Fig. 1, after five years of running,
the region with $\sin^2 2\theta_{e\mu}>0.02$ will be explored.  Thus
this experiment can be used to explore a series of hypotheses.
If a signal is observed,  then the result may lie in the Global Fit allowed
region, consistent with all of the allowed anomalies interpreted
within  a sterile neutrino model.    Alternatively, 
the signal may be inconsistent with the Global Fit hypotheses,
but lie within the Global $\bar \nu_e$ Appearance allowed region,
indicating that the Reactor Anomaly is not due to oscillations, but
that the LSND and MiniBooNE signals do arise from oscillations.   Lastly, if no signal
is observed, then 
according to the argument at the start of the report,   $\sin^2
2\theta_{e\mu}>0.02$ 
can be excluded in all models that respect $CPT$.  In
Fig.~\ref{measure}, we present the measurement capability for five years
of IsoDAR@KamLAND for various true values of the oscillation 
parameters.

\section{IsoDAR@JUNO}

\begin{figure}[t]\begin{center}
{\includegraphics[width=3.2in]{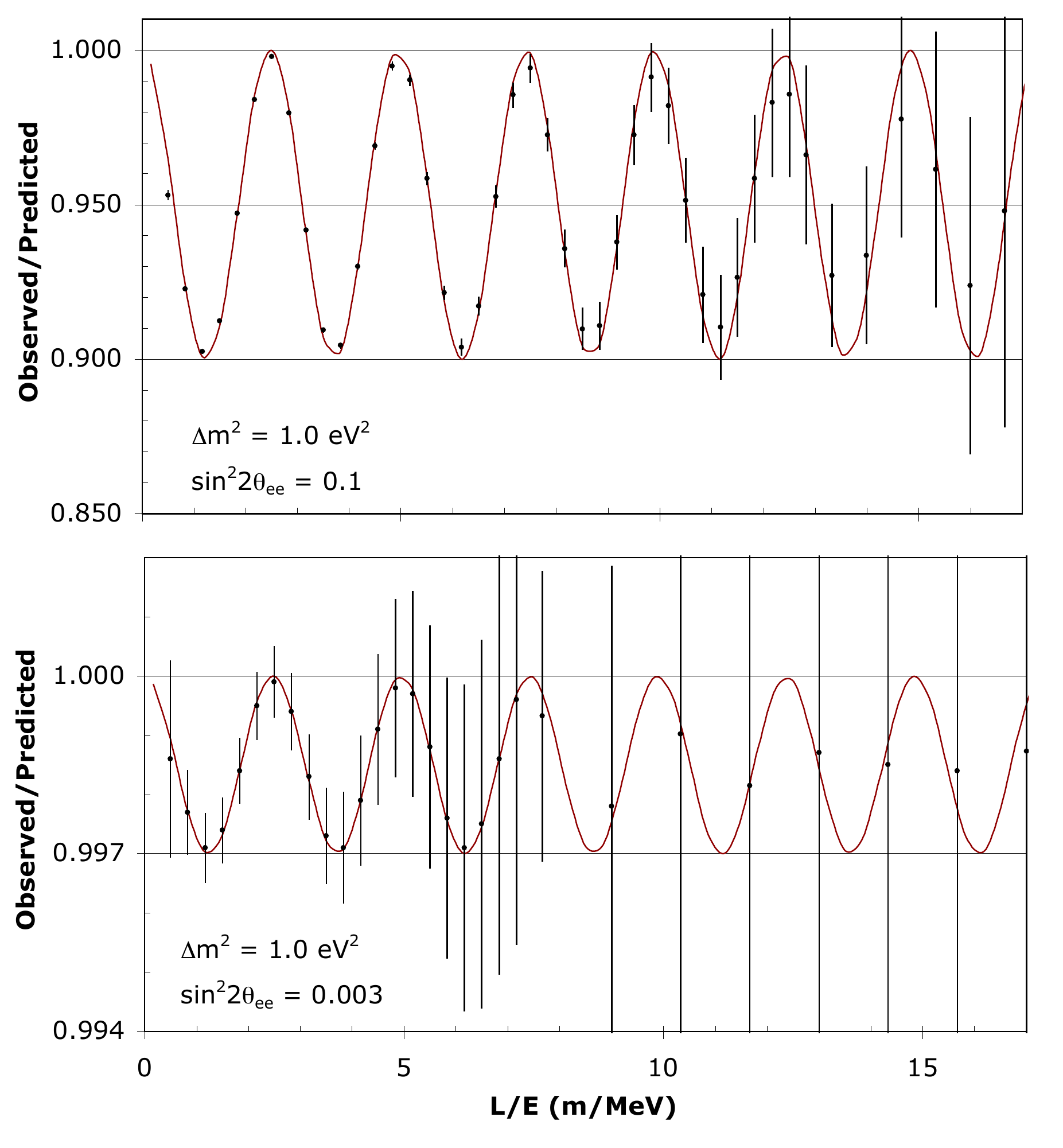}
} 
\end{center}
\vspace{-0.25in}
\caption{The L/E dependence of the oscillation signature after five years 
of IsoDAR@JUNO running for $\Delta m^2 = 1.0 \ {\rm eV}^2$ and $\sin^2
2\theta = 0.1$ (top) --a solution
within the Global Fit allowed region-- and $\sin^2
2\theta = 0.003$ (bottom) --a solution within the Global $\bar \nu_e$ Appearance allowed region.
The black points are the simulated data and the 
solid curve is the oscillation probability with no smearing in the reconstruction
of position and energy.
\label{JUNO_Wave} }
\end{figure}

The IsoDAR source can also be paired with the 20 kton JUNO (formerly, Daya Bay
II) detector.  This is a liquid scintillator detector proposed for a
reactor-based determination of the mass hierarchy \cite{JUNO}.
To calculate the sensitivity, we repeat the analysis of IsoDAR@KamLAND
described in the previous section, assuming that an 
IsoDAR antineutrino source
is run with this much larger and more precise experiment.

The design of JUNO remains under discussion.  For this analysis, we 
used the following parameters.  The active target consists of an
acrylic sphere of 34.5 m in diameter.  The resulting fiducial volume
is 20 ktons.    The PMTs are located at a diameter of 37.5 m.  Beyond
this is a 1.5 m veto region.   This is assumed to be surrounded by
rock.  The depth is expected to be similar to that of KamLAND, 
$\sim 2000$ m.w.e.
The JUNO detector is being carefully designed to achieve excellent 
energy resolution of $3\%/\sqrt{E}$.   We assume that the vertex
resolution is the same as for KamLAND.
The source requires substantial iron and concrete shielding to contain the neutrons that
escape the $^7$Li sleeve, which limits the proximity of IsoDAR to any
large detector.  For these studies, we assume a five-year run with an IsoDAR 
cyclotron source that accelerates deuterons.  The deuteron option gives an enhanced
rate ($\times 2.7$) of antineutrino production as compared to the proton
option described in Ref.~\cite{PRL} and also has a smaller size.
For this analysis, we assume that
the center of the source is located 5.0 m from the edge of the active
region of JUNO.     
This leads to $27.5\times 10^6$ IBD events reconstructed
in the JUNO detector, assuming no oscillation.  

Fig.~\ref{JUNO_sense} shows the sensitivity curve for IsoDAR@JUNO (blue line),
which completely covers the ``Global $\bar\nu_e$ Appearance'' region at 
greater than the 5$\sigma$ CL.  If no oscillation signal is observed,
then all present anomalies, including 
LSND, will be excluded at $5\sigma$ as an indication of neutrino
oscillations.   Thus, this fits the call to decisively address all of
the present high $\Delta m^2$ anomalies.

On the other hand, if an oscillation signal is observed, then precision 
measurements of the L/E dependence
will allow the oscillations to be studied and quantified as shown in
Fig.~\ref{JUNO_Wave}.    The upper plot shows the oscillation signal
for a point in the Global Fit allowed region, which would fit models
if all of the present anomalies are verified as oscillations.  The lower plot shows
the signal for -a solution within the Global $\bar \nu_e$ Appearance
allowed region, which woud fit models if only the LSND-MiniBooNE (antineutrino)
anomalies are verified.

\section{Conclusion}

In this paper, we have shown that IsoDAR@JUNO $\bar \nu_e$
disappearance experiment has sensitivity to cover
the entire short-baseline appearance allowed region at 5$\sigma$.  This allows for a
decisive test of the question of whether the LSND signal arises from
oscillations. We also show that IsoDAR@KamLAND can address 
a substantial region of the allowed space.  These are elegant experiments
because the only assumption is that $CPT$ is a valid
symmetry.

\begin{center}
{ \textbf{Acknowledgments}}
\end{center}

The authors thank the National Science Foundation for support.   We
thank the DAE$\delta$ALUS members and B. Kayser for useful discussion. And we
thank C. Ignarra for providing the global $\bar \nu_\mu
\rightarrow \bar \nu_e$ allowed region in Fig.~\ref{JUNO_sense}.

\end{document}